\documentclass{iopart}
\usepackage{iopams}
\usepackage[colorlinks=true,linkcolor=blue,urlcolor=blue,citecolor=blue]{hyperref}
\begin{document}

\title{Potfit: effective potentials from ab-initio data}

\author{Peter Brommer and Franz G\"ahler}

\address{Institut f\"ur Theoretische und Angewandte Physik (ITAP),
  Universit\"at Stuttgart, Pfaffenwaldring 57, 70550 Stuttgart, Germany}
\ead{\href{mailto:p.brommer@itap.physik.uni-stuttgart.de}{p.brommer@itap.physik.uni-stuttgart.de}}

\begin{abstract}
  We present a program called \emph{potfit} which generates an effective 
  atomic interaction potential by matching it to a set of reference 
  data computed in first-principles calculations. It thus allows to
  perform large-scale atomistic simulations of materials with physically 
  justified potentials. We describe the fundamental principles behind 
  the program, emphasizing its flexibility in adapting to different 
  systems and potential models, while also discussing its limitations.
  The program has been used successfully in creating effective potentials 
  for a number of complex intermetallic alloys, notably quasicrystals.
  \medskip
%

\noindent\emph{Modelling Simulation Mater.\ Sci.\ Eng.\/}  {\bf 15} (2007),
295--304\\online at \url{http://stacks.iop.org/ms/15/295}\\doi:
\href{http://dx.doi.org/10.1088/0965-0393/15/3/008}{10.1088/0965-0393/15/3/008} 
\end{abstract}



\section{Introduction}

Classical effective potentials reduce the quantum-mechanical interactions 
of electrons and nuclei in a solid to an effective interaction between 
atom cores. This greatly reduces the computational effort in molecular 
dynamics (MD) simulations. Whereas first principles simulations are 
limited to a few hundred atoms at most, classical MD calculations with many 
millions of atoms are routinely performed. Such system sizes are possible, 
because molecular dynamics with short-range interactions scales linearly 
with the number of atoms. Moreover, it can easily be parallelized using a
geometrical domain decomposition scheme \cite{itapdb:Allen1987,%
itapdb:Beazley1995b}, thereby achieving linear scaling also in the number 
of CPUs. 
 
The study of many problems in materials science and nanotechnology
indeed requires simulations of systems with millions of atoms.
Quite generally, this is the case whenever long-range mechanical 
stresses are involved. Examples of such problems are the study of 
fracture propagation \cite{itapdb:Roesch2005}, nano-indentation,
or the motion and pinning of dislocations. 
Other problems may be simulated with more moderate numbers of atoms, 
but require very long simulated times, of the order of nanoseconds,
an example of which is the study of atomic diffusion \cite{itapdb:Hocker2006}.
In either case, if large systems and/or long time scales are required,
classical effective potentials are the only way to make molecular
dynamics simulations possible.

The reliability and predictive power of classical MD simulations depend
cruicially on the quality of the effective potentials employed. In
the case of elementary solids, such potentials are usually obtained by 
adjusting a few potential parameters to optimally reproduce a set of reference 
data, which typically includes a number of experimental values like 
lattice constants, cohesive energies, or elastic constants, sometimes 
supplemented with ab-initio cohesive energies and 
stresses \cite{itapdb:Daw1993,itapdb:Chantasiriwan1996}.
In the case of more complex systems with a large variety of local 
environments and many potential parameters to be determined, such an
approach cannot help, however; there is simply not enough reference 
data available.

The force matching method \cite{itapdb:Ercolessi1994} provides a way to
construct physically justified potentials even under such circumstances.  
The idea is to compute forces and energies from first principles for a 
suitable selection of small reference systems and to adjust the 
parameters of the potential to optimally reproduce them.

For that purpose, we developed a program called \emph{potfit}\footnote{
\texttt{\url{http://www.itap.physik.uni-stuttgart.de/\%7Eimd/potfit}}}.
By separating the process of optimization from the form of
the potential, \emph{potfit} allows for maximal flexibility in the choice of
potential model and parametrization. 

The underlying algorithms are described in section~\ref{sec:algo}. Section
\ref{sec:implement} focuses on the implementation of the algorithms,
followed by details on employing \emph{potfit} in section~\ref{sec:results}.
We discuss advantages and limitations of the force matching method and our
implementation in section~\ref{sec:discussion}, and present our conclusions 
in the final section~\ref{sec:conclusion}.


\section{Algorithms}
\label{sec:algo}

As mentioned above, \emph{potfit} consists of two separate parts. 
The first one implements a particular parametrized potential model and 
calculates from a set of potential parameters $\xi_{i}$ the target function 
that quantifies the deviations of the forces, stresses and energies from the
reference values. Wrapped around is a second, potential independent part
which implements a least squares minimization module. As this part is
completely independent of the potential model and just deals with the 
list of parameters $\xi_{i}$, it is fairly straightforward to change the
parametrization of the potential (tabulated or analytic), or even to 
switch to a different potential model.


\subsection{Optimization}

From a mathematical point of view, force matching is a basic optimization
problem: There is a set of parameters $\xi_{i}$, a set of values
$b_{k}(\xi_{i})$ depending on them, and a set of reference values
$b_{0,k}$ which the $b_{k}$ have to match. This leads to the
well-known method of least squares, where one tries to minimize the sum of
squares of the deviations between the $b_{k}$ and the $b_{0,k}$. In our case,
the reference values can either be the components of the force vector
$\vec{f}_{0,j}$ acting on each individual atom $j$, or global data $A_{0,k}$
like stresses, energies, or certain external constraints. We found it
helpful to measure the relative rather than the absolute deviations from 
the reference data, except for very small reference values. The least squares
target function thus becomes
\begin{eqnarray}
  \label{eq:target}
  Z=Z_{\rm{F}}+Z_{\rm{C}},\qquad \\ \mbox{with} \quad
Z_{\rm{F}}=\sum\limits_{j=1}^{N_{A}}\sum\limits_{\alpha=x,y,z}W_{j}
\frac{\left(f_{j_{\alpha}}-f_{0,j_{\alpha}}\right)^{2}}
{\vec{f}_{0,j}^{2}+\varepsilon_{j}},\\\mbox{and} \quad
Z_{\rm{C}}=\sum\limits_{k=1}^{N_{c}}W_{k}
\frac{\left(A_{k}-A_{0,k}\right)^{2}}{A_{0,k}^{2}+\varepsilon_{k}},
\end{eqnarray}
where $Z_{F}$ represents the contributions of the forces, and $Z_{C}$ that 
of the global data. The (small and positive) $\varepsilon_{\ell}$ 
impose a lower
bound on the denominators, thereby avoiding a too
accurate fitting of small quantities which are actually not known to
such a precision. The $W_{\ell}$ are the weights of the different terms.  
It proves useful for the fitting to give the total stresses and the 
cohesion energies an increased weight, although in principle they should 
be reproduced correctly already from the forces. Even if all forces 
are matched with a small deviation only, those deviations can add up 
in an unfortunate way when determining stresses, thus leading to 
potentials giving wrong elastic constants. Including global quantities 
in the fit with a sufficiently high weight supresses such undesired 
behaviour of the fitting process.

As the evaluation of the highly nonlinear target function \eref{eq:target} 
is computationally rather expensive, a careful choice of the 
minimization method has to be made. We chose a combination of 
a conjugate-gradient-like deterministic algorithm \cite{itapdb:Powell1965}
and a stochastic simulated annealing algorithm \cite{itapdb:Corana1987}.

For the deterministic algorithm we take the one described by 
Powell \cite{itapdb:Powell1965}, which takes advantage of the form of the 
target function (which is a sum of squares). By re-using data obtained in
previous function calls it arrives at the minimum faster than standard least
squares algorithms. It also does not require any knowledge of the gradient of
the target function. The algorithm first determines the gradient matrix at the
starting point in the high-dimensional parameter space by finite differences.
The gradient matrix is assumed to be slowly varying around the starting point.
A new optimal search direction towards the minimum is determined by the method
of conjugate gradients. Then, the target function is minimized along this 
direction. This operation is called line minimization. When the minimum is 
found, the direction unit vector replaces one of the basis vectors spanning 
the parameter space. The gradient matrix is updated only with respect to 
this new direction, using the finite differences calculated in the 
line minimization. In this way, no finite differences have 
to be calculated explicitly except in the very first step.  The 
line minimization is performed by Brent's algorithm 
\cite{itapdb:Brent1973} in an implementation taken from the 
GNU Scientific Library \cite{itapdb:Galassi2005}.

The algorithm is restarted (including a calculation of the full gradient 
matrix) when either a step has been too large to maintain the assumption
of a constant gradient matrix, the basis vectors spanning the parameter 
space become almost linearly dependent, or the linear equation
involved in Powell's algorithm cannot be solved with satisfactory numerical
precision.

The other minimization method implemented is a simulated
annealing\cite{itapdb:Kirkpatrick1983} algorithm proposed by 
Corana \cite{itapdb:Corana1987}. While the deterministic algorithm mentioned
above will always find the closest \emph{local} minimum, simulated annealing 
samples a larger part of the parameter space and thus has a chance to end 
up in a better minimum. The price to pay is a computational burdon
which can be several orders of magnitude larger.

For the basic Monte Carlo move, we chose adding Gaussian-shaped bumps to 
the potential functions. The bump heights are normally distributed around 
zero, with a standard deviation adjusted so that on average half of the 
Monte Carlo steps are accepted. This assures optimal progress: Neither are 
too many calculations wasted because the changes are too large to be 
accepted, nor are the steps too small to make rapid progress.


\subsection{Potential models and parametrizations}

The simplest effective potential is a pair potential, which only depends 
on interatomic distances. It takes the form
\begin{equation}
  \label{eq:pairpot}
  V=\sum_{i,j<i}^{N}\phi_{s_{i}s_{j}}(r_{ij}),
\end{equation}
where $r_{ij}$ is the distance between atoms $i$ and $j$, and 
$\phi_{s_{i}s_{j}}$ is a potential function depending on the two atom
types $s_{i}$ and $s_{j}$. This function can either be given in analytic 
form, using a small number of free parameters, like for a Lennard-Jones 
potential, or in tabulated form together with an interpolation scheme for
distances between the tabulation points. Whereas the parameters of an
analytic potential can often be given a physical meaning, such an 
interpretation is usually not possible for tabulated potentials. On the 
other hand, an inappropriate form of an analytic potential may severely 
constrain the optimization, leading to a poor fit. 
For this reason, we chose the functions $\phi$ to be defined by tabulated 
values and spline interpolation, thus avoiding any bias introduced by an 
analytic potential. This choice results in a relatively high number of 
potential parameters, compared to an analytic description of the potentials.
This is not too big a problem, however. Force matching provides enough 
reference data to fit even a large number of parameters. The potential 
functions $\phi$ only need to be defined at pair distances $r$ between 
a minimal distance $r_{\rm{min}}$ and a cutoff radius $r_{\rm{cut}}$, 
where the function should go to zero smoothly.

We found pair potentials to be insufficient for the simulation of 
complex metallic alloys. More suited are EAM (Embedded Atom Method 
\cite{itapdb:Daw1984,itapdb:Daw1993}) potentials, also known as
glue potentials \cite{itapdb:Ercolessi1988}, which have
many advantages over pair potentials in the description of 
metals \cite{itapdb:Ercolessi1988}.

EAM potentials include a many-body term depending on a
local density $n_{i}$:
 
\begin{equation}
  \label{eq:gluepot}
  V=\sum_{i,j<i}\phi_{s_{i}s_{j}}(r_{ij})+
  \sum_{i}U_{s_{i}}
  (n_{i})\qquad
\mbox{with}\quad n_{i}=\sum_{j\neq i}\rho_{s_{j}}(r_{ij}).
\end{equation}
$n_{i}$ is a sum of contributions from the neighbours through a transfer
function $\rho_{s_{j}}$, and $U_{s_{i}}$ is the embedding function that yields
the energy associated with placing atom $i$ at a density $n_{i}$. Again, all
functions are specified by their values at a number of sampling points.

The parameters $\xi_{i}$ specifying a tabulated potential are naturally
the values at the sampling points. Due to the nature of spline interpolation,
either the gradient or the curvature at the exterior sampling points of each
function can also be chosen freely. Depending on the type of potential one
can keep the gradients fixed, or adapt them dynamically by adding them to the
set of parameters $\xi_{i}$.

The EAM potential described by \eref{eq:gluepot} has two gauge degrees of
freedom, i.e., two sets of parameter changes which do not alter the physics 
of the potential:
\begin{equation}
  \label{eq:invarb}
\eqalign{
\rho_{s}(r) \rightarrow \kappa \rho_{s}(r), \cr
U_{s_{i}}(n_{i})  \rightarrow U_{s_{i}}(\case{n_{i}}{\kappa}),}
\end{equation}
and
\begin{equation}
\label{eq:invarmult}
  \eqalign{\phi_{s_{i}s_{j}}(r)\rightarrow
  \phi_{s_{i}s_{j}}(r)+\lambda_{s_{i}}\rho_{s_{j}}(r)+
  \lambda_{s_{j}}\rho_{s_{i}}(r), \cr 
 U_{s_{i}}(n_{i}) \rightarrow U_{s_{i}}(n_{i}) - \lambda_{s_{i}} n_{i}.}
\end{equation}
According to \eref{eq:invarb}, the units of the density $n_{i}$ can be chosen
arbitrarily. We use this degree of freedom to set the units such that the
densities $n_{i}$ computed for the reference configurations are contained
in the interval $(-1;1]$, but not in any significantly smaller interval. 
The transformation \eref{eq:invarmult} 
states that certain energy contributions can be moved freely between the 
pair and the embedding term. An embedding function $U$ which is linear 
in the density $n$ can be gauged away completely. This also makes any 
separate interpretation of the pair potential part and the embedding 
term void; the two must only be judged together. The latter degeneracy 
is usually lifted by choosing the gradients of the $U_{i}(n_{i})$ to vanish 
at the average density for each atom type. \emph{potfit} also uses this 
convention when exporting potentials for plotting and MD simulation. 
As the average density might change during minimization, \emph{potfit} 
internally uses a slightly different gauge: It requires that the 
gradient vanishes at the center of the domain of the respective embedding
function.

\emph{potfit} can perform the transformations
(\ref{eq:invarb},\ref{eq:invarmult}) periodically on its own, thus
eliminiating the need to fix the gauge by an additional 
term in the target function \eref{eq:target}. Unfortunately, for
tabulated functions the transformations cannot be performed exactly 
due to the nature of spline interpolation. A change of gauge therefore
can lead to an increase of the target function, which is why we suppress 
such gauge transformations in the very late stages of a minimization.


\section{Implementation}
\label{sec:implement}

\emph{potfit} is implemented in ANSI C. While the user may specify most 
options in a parameter file read when running the program, some fundamental
choices must be made at compile time, like for example the potential model
used, or whether to allow for automatic gauge transformations in EAM
potentials.  This is a compromise between convenience and computation speed.
Compile time options can be selected by passing them to the make command, 
and thus do not require any changes of the source files. For solving the
linear equations in Powell's minimization algorithm, \emph{potfit}
makes use of routines from the LAPACK library \cite{itapdb:Anderson1999},
which must be installed separately, probably together with the
BLAS library \cite{itapdb:Lawson1979} LAPACK is based on.


\subsection{Parallelization and optimization}

The program spends almost all CPU time in calculating the forces for a 
given potential; finding a new potential to be tested against the reference 
data takes only a tiny fraction of that time. Thus, the only way to
improve performance is to reduce the total time needed for the force
computations, either by minimizing their number, or by making each
force computation faster. Powell's algorithm leaves only little room 
for further reduction of the number of force evaluations. One could 
for instance adjust the precision required in a line minimization.
If the tolerance is too small, time is wasted in refining a 
minimum beyond need, whereas an insufficent precision may stop too far 
from the minimum, thus requiring more steps in total. The choice of
this tolerance was made empirically.

Much more time can be saved by parallelizing the calculation of forces, 
energies, and stresses for a given potential. This is done in a
straightforward way: As the forces, energies, and stresses of
the different reference configurations can be computed independently,
we simply distribute the reference configurations on several processes.
Before the force computation, the potential parameters are distributed
to all processes, and afterwards the computed forces, energies and
stresses are collected. The communication is performed using the
standard Message Passing Interface (MPI \cite{itapdb:Gropp1999}).
This simple parallelization scheme works well as long as the number 
of configurations per process does not drop below 10 to 15. 
Otherwise, the communication overhead starts to show up, 
and load balancing problems may appear.
A shared memory OpenMP parallelization also exists, 
but produces inferior results. 

In force matching, the reference configurations stay fixed. Therefore,
all distances between atoms remain fixed, and \emph{potfit} can use 
neighbour lists, which need to be computed only once at startup. 
In fact, for each neighbour pair all data required for spline
interpolation are pre-computed, allowing for a fast lookup of
the tabulated functions. This data needs to be recomputed only
when the tabulation points of a function are changed.


\subsection{Input and output files}

Tabulated potential functions can be specified with equidistant
or with arbitrary tabulation points. For equidistant tabulation 
points, the boundaries of the domain and the number of sampling 
points of each function are read from the potential file, followed 
by a list of function values at the sampling points and the 
gradients at the domain boundaries. In the case of free tabulation 
points, only their number is specified at the beginning of the 
potential file, followed by a list of argument-value pairs 
and again the gradients of the potential functions at the domain 
boundaries.

Reference configuration files contain the number of atoms, the box vectors,
the cohesive energy, and the stresses on the unit cell, followed by a list 
of atoms, with atom species, position and reference force for each atom.
Such reference configuration files can simply be concatenated.

\emph{potfit} was designed to cooperate closely with the first-principles
code VASP \cite{itapdb:Kresse1993,itapdb:Kresse1996} and with 
IMD \cite{itapdb:Stadler1997a}, our own classical MD code.
VASP, which is a plane wave code implementing ultrasoft pseudopotentials
and the Projector-Augmented Wave (PAW) 
method \cite{itapdb:Blochl1994,itapdb:Kresse1999}, is used to compute
the reference data for the force matching, whereas the resulting
potentials are intended to be used with IMD. For this reason,
\emph{potfit} provides import and export filters for potentials and 
configurations to communicate with these programs. These filters are 
implemented as scripts, which can easily be modified to interface
with other programs.


\section{Results and validation}
\label{sec:results}

As a first test, \emph{potfit} should be able to recover a classical 
potential from reference data computed with that potential. For this 
test, we used snapshots from several molecular dynamics runs as reference 
structures, first for a Lennard-Jones fcc solid, then for a complex 
Ni-Al alloy simulated with EAM potentials \cite{itapdb:Ludwig1995a}.
In order to ensure that all reference data presented to \emph{potfit} 
is consistent, the potentials were approximated by cubic spline 
polynomials, in the same way as \emph{potfit} represents the potentials.
With such reference data and starting with vanishing potential 
functions, \emph{potfit} could in both cases perfectly recover the 
potentials. This test therefore demonstrates the correctness 
of the program. One should keep in mind, however, that reference 
data from ab-inito computations often cannot be reproduced perfectly
by any classical potential.

Our primary research interest are quasicrystals \cite{itapdb:Trebin2003} 
and other complex metal alloys, for which good potentials are
hardly available. \emph{potfit} has been developed in order to
generate effective potentials for such complex metal alloys, 
which feature large (or even infinite) unit cells, several atom 
species, and a wide variety of different local environments. So 
far, force matching had been used mainly to determine potentials
for monoatomic metals and a small selection of relatively simple 
binary alloys.

As a first application beyond simple alloys, we have developed 
potentials for the quasicrystalline and nearby crystalline phases 
in the systems Al-Ni-Co, Ca-Cd, and Mg-Zn. Due to the complexity
of the structures and also due to the choice of tabulated potential
functions, a relatively large number of potential parameters
is required. This is especially true for ternary EAM potentials,
which comprise 12 tabulated functions, with 10--15 tabulation
points each. Correspondingly, a relatively large amount of reference 
data is required. A computationally 
efficient implementation of the force matching method is therefore
essential. It turned out that \emph{potfit} scales well under those
circumstances and is up to its task.

Although the potentials to be generated are intended for (aperiodic)
quasicrystals and crystals with large unit cells, all reference
structures have to be periodic crystals with unit cell sizes suitable 
for the ab-initio computation of the reference data. On the 
other hand, the reference structures should approximate the quasicrystal
in the sense, that all their unit cells together accommodate all relevant
structural motifs. To do so, they must be large enough. For instance, 
the quasicrystalline and related crystalline phases of Ca-Cd and Mg-Zn 
consist of packings of large icosahedral clusters in different 
arrangements. Reference structures must be able to accomodate such 
clusters. A further constraint is, that the unit cell diameter must be 
larger than the range of potentials. We found that reference structures 
with 80--200 atoms represent a good compromise between these requirements.

Starting from a selection of basic reference structures, further
ones were obtained by taking snapshots of MD simulations with model
potentials at various temperatures and pressures. Also samples which 
were strained in different ways were included. For all these reference 
structures, the ab-initio forces, stresses and energies werde 
determined with VASP, and a potential was fitted to reproduce these 
data. As reference energy, the cohesive energy was used, i.e., the
energies of the constituent atoms was subtracted from the VASP 
energies. Instead of absolute cohesive energies one can also use the 
energy relative to some reference structure. Once a first version of 
the fitted potential was available, the MD snapshots were replaced 
or complemented with better ones obtained with the new potential, 
and the procedure was iterated. 

As expected, no potential could be found which would reproduce the
reference data exactly. During the optimization, the target function 
(\ref{eq:target}) does not converge to zero, which indicates that
quantum mechanical reality (taking density functional theory as reality)
is not represented perfectly by the potential model used. 
The forces computed from the optimal potential typically differ 
by about 10\% from the reference forces, which seems acceptable.
For the energies and stresses a much higher agreement could be
reached. Cohesion energy differences for instance can be reproduced 
with an accuracy better than 1\%.

The generated potentials were then used in molecular dynamics 
simulations to determine various material properties, such as the
melting temperature and the elastic constants, for which values 
consistent with experiment were obtained. The Ca-Cd potentials
were especially tuned towards ground-state like structures,
whose energies are reproduced with high accuracy, in agreement
with ab-initio results. Details of these applications can be found 
in \cite{itapdb:Hocker2006,itapdb:Brommer2006,itapdb:Mihalkovic2006,%
itapdb:Brommer2007}. 
Probably the best tested EAM potential constructed with \emph{potfit}
was obtained by R\"osch, Trebin and Gumbsch \cite{itapdb:Roesch2006}.
This potential is intended for the simulation of crack propagation
in the C15 Laves Phase of NbCr$_{2}$, and has undergone a broad 
validation. These authors calculated the lattice constant, the elastic 
constants and the melting temperature and compared these values to 
experimental and ab-initio results with reasonable success. They 
also studied relaxation of surface atoms, surface energy and the 
crack propagation in NbCr$_{2}$. According to the authors 
\cite{itapdb:Roesch2006}, the force-matched potentials created with 
\emph{potfit} clearly outperform previously published potentials.
But this example also shows \cite{itapdb:Roesch2006} that a large 
number of fitting-validation cycles are usually required, before
a usable and satisfactory potential is obtained. This makes force 
matching a time-consuming and tedious process.


\section{Discussion}
\label{sec:discussion}

\subsection{Transferability} It should be kept in mind that 
force-matched potentials will only work well in situations they have
been trained to. Therefore, all local environments that might occur in the 
simulation should also be present in the set of reference configurations.
Otherwise the results may not be reliable. Using a very broad selection
of reference configurations will make the potential more {\em transferable},
making it usable for many different situations, e.g. for different
phases of a given alloy. On the other hand, giving up some transferability
may lead to a higher precision in special situations. By carefully 
constraining the variety of reference structures  
one may generate a potential that 
is much more precise in a specific situation than a general purpose 
potential, which was trained on a broader set of reference structures.
The latter potential, on the other hand, will be more versatile, but less
accurate on average. 
Finding sufficiently many suitable reference structures might not
always be trivial. For certain complex structure like quasicrystals,
there may be only very few (if any) approximating periodic structures 
with small enough unit cells.



\subsection{Optimal number and location of sampling points}
Each reference database has an optimum number of parameters it can support.
Using too few parameters, the potential functions lack flexibility. On the
other hand, exceeding this number may lead to overfitting beyond the limit of
the potential model. \emph{potfit} cannot determine that optimal number
automatically, but there is a simple strategy the user can employ. The set of
reference configurations is split in two subsets, one of which is used for
fitting and the other for testing the potential. If the root-mean-square (rms)
deviation of the test set significantly exceeds that of the fitting set, the
database is probably overfitted \cite{itapdb:Robertson1993}. By starting with
a relatively low number of parameters, that is increased as long as the rms of
the \emph{testing} stage decreases, one can arrive at the optimal number of
parameters \cite{itapdb:Mishin1999}.

This strategy also helps in dealing with oscillatory artefacts of the spline
interpolation: If the sampling points are not spaced too densely, and there is
enough data to support each tabulation point, artificial wiggles are 
suppressed. \emph{potfit} provides the frequency with which each 
tabulation interval is accessed during an evaluation of the target 
function \eref{eq:target}. With this information, sampling point density 
can be reduced for distances that do not appear frequently enough in the 
reference configurations.



\subsection{Number of atom types and choice of reference structures} 

The most obvious impact of an increasing number of atom types
is the corresponding increase in the number of potential parameters.
For instance, an EAM potential for $n$ atom types requires
$n(n+1)/2+2n$ tabulated functions, each with 10 to 15 tabulation
points. For a ternary system, this already amounts to the order
of 150 potential parameters. Whereas such a number of parameters
can still be handled, an increasing number of atom types leads to 
yet another problem, which is more serious. To see this, it must be 
kept in mind that any potential function depending on the interatomic 
distance must be determined for the entire argument range between 
$r_{\rm{min}}$ and $r_{\rm{cut}}$. If tabulated functions are used, 
for each tabulation interval there must be distances actually occuring
in the reference structures, for otherwise there are potential 
parameters which do not affect the target function, and which 
conseqently cannot be determined in the fit. The requirement that 
all distances for all combinations of atom types actually occur 
in the reference structures becomes especially problematic if 
the atoms of one type form only a small minority, in which case 
some distances between such atoms might be completely absent in 
all reasonable reference structures. If the number of atom types 
is large, there is unfortunately always at least one element which 
is a minority constituent. In such situations it might be 
unavoidable to use a much broader selection of reference 
structures with varying stoichiometry, instead of a fixed 
stoichiometry with a minority constituent. It might even be 
necessary to include energetically less favourable configurations
to provide a complete set of reference data.

Another solution would be to use a non-local (or less local) 
parametrisation of the potential functions, like a superposition 
of broad gaussians or functions given by analytic formulae. 
Changing one parameter can then affect the function over a 
broader range of arguments, making it again possible to fit the 
function even if only sparse information on it is provided by the
reference data.
Potentials represented in this way would also not suffer from 
the wiggle artefacts of spline interpolation described above. 


\subsection{Experimental values as reference data}

\emph{potfit} does currently not use experimental data during force 
matching. The potentials are determined exclusively from ab-initio 
data, which means they cannot exceed the accuracy of the first principles 
calculations. While it is possible, in principle, to support also the
comparison to experimental values, we decided against such an addition. 
For once, available experimental values can often not be calculated directly
from the potentials, so determining them would considerably slow down the
target function \eref{eq:target} evaluation. Secondly, experimental 
values often also depend on the exact structure of the system, which in 
most cases is not completely known beforehand for complex structures,
for instance due to fractional occupancies in the experimentally
determined structure model. A better way to use experimental data is to 
test whether the newly generated potentials lead to structures that 
under MD simulation show the behaviour known from experiment.


\section{Conclusion}
\label{sec:conclusion}

Large scale molecular dynamics simulations are possible only with classical
effective potentials, but for many complex systems physically justified 
potentials do not exist so far. Our program \emph{potfit} allows the 
generation of effective potentials even for complex binary and ternary 
intermetallics, adjusting them to ab-initio determined reference data 
using the force matching method. Potentials for several complex 
intermetallic compounds have been generated, and were successfully 
used in molecular dynamics studies of various 
properties \cite{itapdb:Hocker2006,itapdb:Brommer2006,itapdb:Mihalkovic2006,%
itapdb:Roesch2006}. 
It should be emphasized, however, that constructing potentials is 
still tedious and time-consuming. Potentials have to be thoroughly
tested against quantities not included in the fit. In this process,
candiate potentials often need to be rejected or refined. Many
iterations of the fitting-validation cycle are usually required. 
It takes experience and skill to decide when a potential is finished 
and ready to be used for production, and for which conditions and 
systems it is suitable. \emph{potfit} is only a tool that assists in 
this process. 
Flexibility and easy extensibility was one of the main design goals
of \emph{potfit}. While at present only pair and EAM potentials with
tabulated potential functions are implemented in \emph{potfit}, 
it would be easy to complement these by other potential models, 
or to add support for differently represented potential functions.

\ack This work was funded by the Deutsche Forschungsgemeinschaft through
Collaborative Research Centre (SFB) 382, project C14. Special thanks go to
Stephen Hocker and Frohmut R\"osch for fruitful discussion and feedback, and
to Hans-Rainer Trebin for supervising the thesis work of the first author.


\section*{References}
\bibliographystyle{iopart-num}
\providecommand{\newblock}{}

\end{document}